\documentclass{article}

\usepackage{arxiv}

\usepackage[utf8]{inputenc} 
\usepackage[T1]{fontenc}   
\usepackage{hyperref}       
\usepackage{url}            
\usepackage{booktabs}       
\usepackage{amsfonts}       
\usepackage{nicefrac}       
\usepackage{microtype}      
\usepackage{graphicx}
\usepackage{natbib}
\usepackage{doi}

\usepackage{siunitx}
\usepackage{caption}
\newcommand{\ra}[1]{\renewcommand{\arraystretch}{#1}}
\usepackage[table]{xcolor}
\usepackage{nicematrix}
\usepackage{lscape} 
\usepackage{lineno}

\title{Quantifying resilience and the risk of regime shifts under strong correlated noise}

\date{\today}

\author{ \href{https://orcid.org/0000-0003-0529-7926}{\includegraphics[scale=0.06]{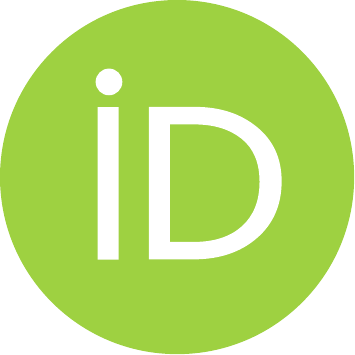}\hspace{1mm}Martin Heßler}\thanks{Center for Nonlinear Science, Westphalian Wilhelms-University Münster, 48149 Münster, Germany} \\
	Institute for Theoretical Physics\\
	Westphalian Wilhelms-University Münster\\
	48149 Münster, North Rhine-Westphalia, Germany\\
	\texttt{m\_{}hess23@uni-muenster.de} \\
	\And
	\href{https://orcid.org/0000-0003-0986-0878 }{\includegraphics[scale=0.06]{orcid.pdf}\hspace{1mm}Oliver Kamps} \\
	Center for Nonlinear Science\\
	Westphalian Wilhelms-University Münster\\ 
	48149 Münster, North Rhine-Westphalia, Germany\\
	\texttt{okamp@uni-muenster.de} \\
}

\renewcommand{\headeright}{\emph{arXiv} Preprint}
\renewcommand{\undertitle}{\emph{arXiv} Preprint}
\renewcommand{\shorttitle}{Quantifying resilience under strong correlated noise}

\hypersetup{
pdftitle={Quantifying resilience and the risk of regime shifts under strong correlated noise},
pdfsubject={resilience, regime shifts, leading indicators, early warning signals},
pdfauthor={Martin Heßler, Oliver Kamps},
pdfkeywords={ecology,regime shift, early warning signals, leading indicator, critical transition},
}

\begin{document}
\maketitle

\begin{abstract}
Early warning indicators often suffer from the shortness and coarse-graining of real-world time series. Furthermore, the typically strong and correlated noise contributions in real applications are severe drawbacks for statistical measures. Even under favourable simulation conditions the measures are of limited capacity due to their qualitative nature and sometimes ambiguous trend-to-noise ratio. In order to solve these shortcomings, we analyse the stability of the system via the slope of the deterministic term of a Langevin equation, which is hypothesized to underlie the system dynamics close to the fixed point. The open-source available method is applied to a previously studied seasonal ecological model under noise levels and correlation scenarios commonly observed in real world data. We compare the results to autocorrelation, standard deviation, skewness and kurtosis as leading indicator candidates by a Bayesian model comparison with a linear and a constant model. We show that the slope of the deterministic term is a promising alternative due to its quantitative nature and high robustness against noise levels and types. The commonly computed indicators apart from the autocorrelation with deseasonalization fail to provide reliable insights into the stability of the system in contrast to a previously performed study in which the standard deviation was found to perform best. In addition, we discuss the significant influence of the seasonal nature of the data to the robust computation of the various indicators, before we determine approximately the minimal amount of data per time window that leads to significant trends for the drift slope estimations.
\end{abstract}


\section{Introduction}\label{sec: introduction}

Even if the idea of universal early warning indicators (\cite{a:dakos12, a:Dakos2009, a:Scheffer2015, a:Liang2017}) for critical transitions is a fascinating and attractive vision throughout the fields of ecology, climate research, biology, power grids (\cite{a:veraart11, a:Drake2010, a:Dakos2017, a:livina2010, a:Livina2015, a:Lenton2012, a:cotilla-sanchez12}), its potential for social and economical sciences (\cite{a:Jusup2022, a:Helbing2014}) and much more (\cite{a:izrailtyan, Chadefaux2014,a:Leemput2013}), the research done over the years in this field has discovered plenty of problems, drawbacks and limitations of the proposed leading indicators (\cite{a:Clements2015, a:Hastings2010, a:Ditlevsen2010, a:Wilkat2019}). The difficulties and limitations result from the sometimes mentioned problematic claim of ``universality'' which is hard or impossible to achieve. Just by definition the mentioned universality is already limited to special cases of regime shifts as bifurcation-induced tipping events (\cite{a:scheffer09, a:Ritchie2017, a:Ashwin2012}), because the leading indicators are a consequence of the commonly observed phenomenon of \textit{critical slowing down} prior to a bifurcation or \textit{flickering} in noisy bistable systems (\cite{a:scheffer12, a:Wissel1984, a:Schroeder2005}). Critical slowing down is the increased relaxation time of perturbations near a bifurcation whereas flickering determines jumps of a system between two alternative stable states. Furthermore, a successful detection of a critical transition depends on the eigen-direction in which the transition takes place and the time series at hand (\cite{a:Boerlijst2013}). Apart from that it remains difficult to get an impression of the leading indicators' quality applied to real world systems because the tests are often performed with historical test data for which is known that a transition is present (\cite{a:Boettiger2012}).\\ 
Following this argumentation it is proposed to design specialised indicators in specific fields of research or systems that are known at least in part (\cite{a:Perretti2012, a:Gsell2016, a:Dablander2020}). One of those research areas is the field of ecology in which standard leading indicators as autocorrelation at a lag of one (AR1), the standard deviation (std) $\hat{\sigma}$, the skewness $\gamma$ or the kurtosis $\omega$ are often very limited in their applicability due to high correlated noise contributions and low sampled short time series that are characteristic because of the limits imposed by the experimental and funding resources as stated in \cite{a:Bissonette1999, a:Perretti2012}. However, due to the rapid developments in sensor and information processing techniques at least the sampling limitations might be partially overcome in future ecological studies e.g. due to deep-learning image recognition techniques of animal-tracking camera or satellite data (\cite{a:Francisco2020, a:Duporge2020,a:Zhao2020}) or acoustic telemetry systems (\cite{a:Aspillaga2021}). Furthermore, even in simulations in which the afore-mentioned practical limitations do not play a role, the inherent design of the indicators raises problems. As discussed in \cite{a:Biggs2009} the standard leading indicator candidates are difficult to interpret because of their qualitative nature: They are designed upon trend changes which can be too gradual and ambiguous to rely on for decision-makers. And in addition, unfortunately these changes are often realized too late for policymakers to adapt management and avoid uprising transitions. Therefore, in the case that a developed early warning measure should be applicable in practise the authors of \cite{a:Biggs2009} claim that it
\begin{quotation}
would rely on: (i) defining critical levels of the regime shift indicators, (ii) linking these critical levels to long-term sustainable impact levels, and (iii) finding or developing indicators that have critical levels that are relatively transferable across different ecosystem types.
\end{quotation}
Based on these demands (\cite{a:Biggs2009}) and the poor performance of standard leading indicator candidates under strong correlated noise found in \cite{a:Perretti2012}, we want to introduce the alternative drift slope estimation \cite{a:Hessler2021, url:GitHessler2021, url:DocsHessler2021} to tackle the problem of anticipating an ecological regime shift and compare it to the above mentioned indicators. Similar to \cite{a:carpenter11} the alternative approach considers the data to be generated by a stochastic differential equation of the Langevin form (\cite{b:Kloeden1992})
\begin{linenomath*}
\begin{equation}
\dot{x}(x,t) = h(x(t),t) + g(x(t), t) \Gamma (t),\label{eq:langevin}
\end{equation}
\end{linenomath*}
where the drift $h(x(t),t)$ captures the deterministic part of the system dynamics under the stochastic influence of a Gaussian and $\delta$-correlated noise process $\Gamma (t)$ that scales with the diffusion $g(x(t), t)$. 
The method estimates the parameterized drift and diffusion terms via Markov Chain Monte Carlo sampling (MCMC) and calculates the drift slope $\zeta$ in the fixed point $x^*$ in rolling windows as a resilience measure of the system. The drift slope $\zeta$ is negative for stable systems and increases with proceeding destabilization. A zero crossing of the drift slope corresponds to a regime shift (\cite{a:Hessler2021}).
In principle, drift and diffusion can also be directly estimated via the Kramers-Moyal expansion (\cite{a:Friedrich1997, a:Friedrich2000}). Nevertheless, this direct drift estimation does not result in stable estimates under the noise conditions of the investigated ecological model (cf. supplementary material (\cite{a:suppleco})). For that reason, we have chosen a fully Bayesian approach, extending the maximum likelihood formulation in \cite{a:Kleinhans2011}, since it allows for a significantly more stable estimation of the slope, includes a straight-forward calculation of credibility bands without approximation by Wilks' theorem and accounts for possibly multi-peaked probability densities.\\
In this study we show that in contrast to the common qualitative leading indicator candidates, the method provides a quantitative and easy-to-interpret resilience measure which is able to fulfill the requirements stated in \cite{a:Biggs2009} for the system discussed there under noise conditions typical for ecological experimental time series data, i.e. correlated strong noise influence (\cite{a:Perretti2012}). In addition, we discuss the important role of the seasonal nature of the data that affects the trend quality of the time series analysis methods. The performance of the early warning signals is tested by comparing the probability that  the data might be explained by a linear trend or a constant model with a Bayesian model comparison. The drift slope $\zeta$ and - if the seasonality is taken into account - the AR1 provide reliable results in our study. Interestingly, in contrast to previous results regarding the same system (\cite{a:Perretti2012}) the AR1 seems to be preferred to the insignificant standard deviation. In this context it is important to mention that the AR1 would be only considered a reliable indicator if both the AR1 and std $\hat{\sigma}$ would increase at the same time.  Apart from this we can reproduce the findings of a generally poor performance of the standard leading indicator candidates (\cite{a:Perretti2012}). In the end, the drift slope seems to be a promising alternative to common leading indicators because of its quantitative nature, easy interpretation and robustness to strong and colored noise contributions. However, its applicability remains limited to situations in which it is possible to generate the necessary amount of data which is around 50 data points per year for the investigated ecological model.\\
The ecological system is presented in section \ref{sec: eco model}. In section \ref{sec: method} we introduce the Bayesian Langevin estimation scheme as well as the methods to estimate the statistical leading indicators in the subsections \ref{subsec: BL estimation} and \ref{subsec: statistical measures}, before we summarize the significance test approach via Bayes factors in subsection \ref{subsec: model comparison}. The results of the applied leading indicators are discussed in section \ref{sec: results} which is divided into three parts: First, the drift slope results are presented in subsection \ref{subsec: drift slope analysis}. Second, the drift slope performance as leading indicator is compared to established candidates via a Bayesian model comparison in subsection \ref{subsec: comparison}, before the needed minimum amount of data per window for the drift slope estimation for the model at hand is defined in subsection \ref{subsec: ws limits}. Finally, we summarize our findings in section \ref{sec: conclusion}.\\
\section{Ecological model}\label{sec: eco model}
\begin{figure}[!htbp]
\includegraphics[width=0.99 \linewidth]{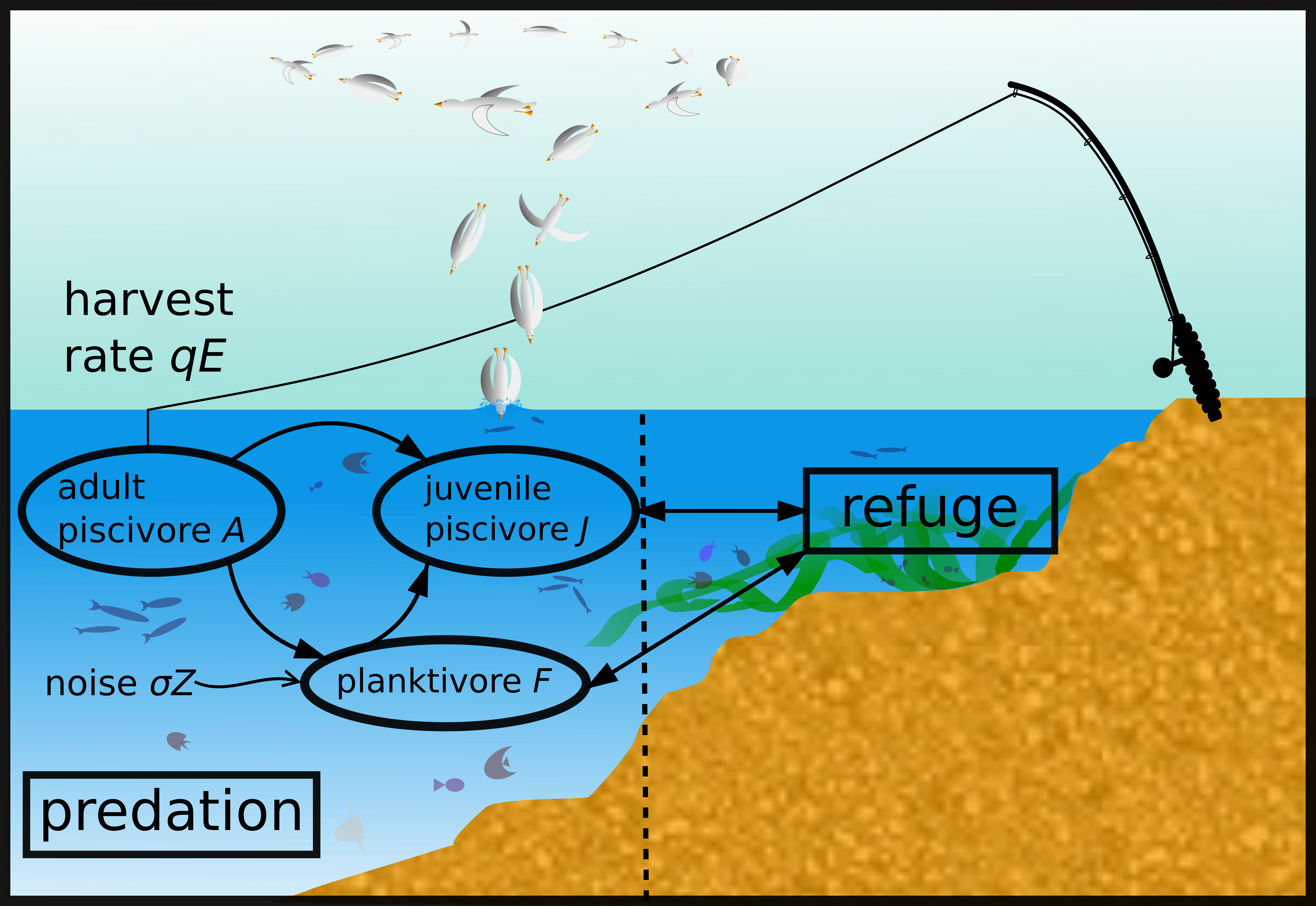}
\caption[Ecological model]{A scheme of the considered foodweb model. In the predation area the adult piscivores $A$ hunt the juvenile piscivores $J$ and the planktivores $F$ which only hunt juvenile piscivores $J$. Both, the juvenile piscivores $J$ and the planktivores $F$ can hide themselves in a refuge area in order to retire. External white or colored stochastic influence $Z$ is added to the planktivore population with the noise level $\sigma$. We discuss the possibility of regime shifts due to high angling pressure represented by the harvest rate $qE$ which is given as the product of catchability $q$ and the effort $E$. Here, the model is restricted to fish, but in general other animals, as e.g. some seabirds, are included in the term ``piscivores''.}
\label{fig: eco model}
\end{figure}
\begin{figure*}[!htbp]
\includegraphics[width=0.98 \linewidth]{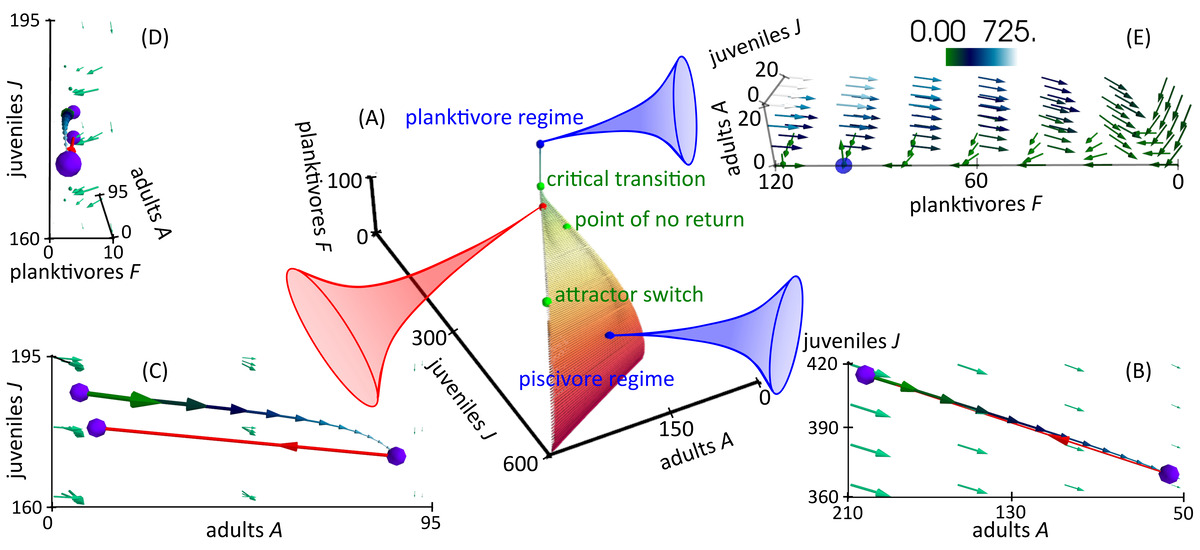}
\caption[Ecological model state space]{The destabilizing journey from the piscivore regime to the planktivore regime is shown in (A) with time color-coded from red to blue. The harvest rate $qE$ is increased linearly to obtain the trajectory. The system starts in the stable piscivore-dominated regime. One realisation is exemplarily shown in (B) for harvest rate $qE = 1.38$. Only a tiny amount of planktivores $0<F<1$ is present and therefore the dynamics are reduced to the $A$-$J$ plane. Juveniles $J$ and adults $A$ of the piscivore species decrease over one year (vector line with time color-coded from green to blue with start and end points given by the violet markers) before the population is updated by the species-intrinsic reproduction rate at the beginning of the new year mirrored by the discrete maturation map (red vector line). The overall surrounding flow is shown as a green vector field with vector lengths reflecting the flow strength. These periodic dynamics are stable until the harvest rate reaches the attractor switch point $qE_{\rm switch} \approx 1.78$. For $qE > qE_{\rm switch}$ the periodic cycle opens up as observable in (C) for an exemplary harvest rate $qE = 2.42$. The discrete maturation step cannot compensate the loss of the piscivore population completely anymore as visible by comparison of the start and end positions marked by violet points of the one-year dynamics which leads to always stronger reduction of the piscivore population over the years. At the same time the planktivore population starts to increase slightly as observable in figure (D) for the same $qE$. The figure is aligned to the planktivore $F$-axis. The green signed critical transition in (A) is subjectively defined for the planktivores $F>21$ for the first time which coincides with the region after which the planktivore species grows drastically. The point of no return around $qE \approx 2.23$ is defined inside of the transition interval between the attractor switch point and the transition point as the point after which even an abrupt reduction of the angling pressure to $qE < 0.1$ cannot avoid the transition to the planktivore-dominated state anymore. At last the new stable state is illustrated in three dimensions in figure (E). The blue point marks the stable planktivore regime with $F \approx 100$, whereas the piscivores are extinct. The flow field strength is color-coded for better resolution.}
\label{fig: state space}
\end{figure*}
In order to investigate the performance of the Bayesian stability analysis tool under rather realistic conditions in the field of ecology the multi-species model derived in \cite{a:Carpenter2004}, described in detail in \cite{a:Biggs2009} and used as a basis of leading indicator performance tests in \cite{a:Perretti2012} is simulated via the Euler-Maruyama scheme. The ecological system consists of three parties: juvenile piscivores ($J$), adult piscivores ($A$) and planktivores ($F$). The model contains a continuous ``monitoring interval'' 
\begin{align}
\frac{\textrm{d} A}{\textrm{d} t} &= -qEA \\
\frac{\textrm{d} F}{\textrm{d} t} &= D_{\textrm{F}}(F_{\textrm{R}} - F) - c_{\textrm{FA}}FA + \sigma Z \\
\frac{\textrm{d} J}{\textrm{d} t} &= - c_{\textrm{JA}}JA - \frac{c_{\textrm{JF}}\nu FJ}{h + \nu + c_{\textrm{JF}}F}
\end{align}
and a discrete annual ``maturation interval'' realized as the map equations
\begin{align}
    A_{y + 1} &= s(A_{y;t = 1} + J_{y;t = 1}) \\
    F_{y + 1} &= F_y \\
    J_{y + 1} &= fA_{y + 1},
\end{align}
where the index $y; t = 1$ means the abundance of each party at the end of the monitoring interval (i.e. $t = 1$) of the corresponding year $y$. In the map $s$ determines the survivorship between maturation intervals and $f$ the fecundity rate of the adult piscivores $A$. The harvest rate of the adult piscivores is determined via the product of the catchability $q$ and the effort $E$. The planktivores exchange between a protected area, the so-called refuge reservoir $F_{\textrm R}$ and the foraging arena $D_{\textrm F}$. The parameters $c_{i,j}$ with $i,j = \lbrace A, F, J\rbrace$ model the consumption or control rates of $i$ by $j$. Besides, the piscivores become vulnerable to planktivores with the rate $\nu$ and enter their refuge with $h$. Environmental stochasticity of the lower level of the food web is incorporated via $Z$. White noise $Z_{\rm white}$ corresponds simply to a Wiener process $Z_{\rm white} = {\textrm d}W$ with spectral power $P_{\rm white} \approx 7.5$. Pink noise $Z_{\rm pink}$ is obtained by the following procedure: 
\begin{enumerate}
    \item Fourier transform $(\mathcal{F}\xi)(f)$ of a white noise signal $\xi(t)$ with $f$ denoting the frequencies,
    \item adjusting the obtained power spectrum by a power law $\sim e^{-\beta_{\rm pink}}$ with $\beta_{\rm pink} = 0.8$ (cf. \cite{a:Perretti2012} for comparability),
    \item and finally an inverse Fourier transform $(\mathcal{F}^{-1}(\mathcal{F}\xi '))(t)$ of the adjusted power spectrum $(\mathcal{F}\xi ')(f)$. The pink noise data is thus $Z_{\rm pink} = (\mathcal{F}^{-1}(\mathcal{F}\xi ')(t)$.
\end{enumerate}
 To keep close to the former studies of \cite{a:Perretti2012} the red noise signal $Z_{\rm red}$ is computed via the Ornstein-Uhlenbeck process
\begin{align}
    {\textrm d} Z_{\rm red} = - \phi Z_{\rm red} {\textrm d} t + \sqrt{2\phi} {\textrm d} W
\end{align}
with $\phi = 0.53$ which results in a spectral exponent $\beta_{\rm red} \approx 1.6$ (cf. \cite{a:Perretti2012} for comparability). The total powers $P$ of the correlated signals are adjusted to be approximately equal $P_{\rm pink}\approx P_{\rm red} \approx 15.4$.
For each of the three noise types the model is evaluated for three different noise intensities, explicitly
\begin{align}
    \sigma {\textrm d} t &= 0.002 \\
    \sigma {\textrm d} t &= 0.044 \\
    \sigma {\textrm d} t &= 0.09
\end{align}
with a time step of ${\textrm d} t = 1/50$.
The realisations of the model are computed with the parameters explicitly listed in table \ref{table: eco model params} and chosen analogously to \cite{a:Perretti2012} apart from the initial harvest rate $qE_{\rm init}$ that is chosen to be $qE_{\rm init} = 1$ instead of $qE_{\rm init} = 1.5$ in order to widen the temporal resolution of the stable regime. The parameter choice in \cite{a:Perretti2012} follows approximately experimentally observed values in ecological systems of that kind, especially for the noise strength (\cite{a:Reed2004}), the noise power law exponents (\cite{a:Steele1985, a:Vasseur2004}) and the increase in angling pressure (\cite{a:Pope1996}). Note that as stated in table \ref{table: eco model params}, the rate of linear destabilization $\Delta (qE)$ as all the other parameters is chosen analogously to \cite{a:Perretti2012} and thus, the choice $qE_{\rm init} = 1$ does not affect the comparability. 
\begin{table*}
\ra{1.4}
    \centering
    \caption{The parameter values of the ecological model with short definitions.}
    \label{table: eco model params}
    \begin{tabular}{@{}lrl}\toprule[1.3pt]
    parameter &  value &  short definition \\
    \midrule
        $qE_{\textrm{init}}$ & $1$ & initial harvest rate \\
        $\Delta (qE)$ & $0.013$ & change of harvest rate per year \\
        $F_{\textrm R}$ & $100$ & refuge reservoir for planktivores \\
        $D_{\textrm F}$ & $0.1$ & foraging arena \\
        $c_{\textrm FA}$ & $0.3$ & rate at which adult piscivores consume planktivores\\
        $c_{\textrm JA}$ & $0.001$ & control of juvenile piscivores by adult piscivores\\
        $c_{\textrm JF}$ & $0.5$ & rate at which planktivores consume juvenile piscivores\\
        $\nu$ & $1$ & rate at which juvenile piscivores become vulnerable against planktivores\\
        $h$ & $8$ & rate at which juvenile planktivores enter the refuge\\
        $f$ & $2$ & fecundity rate of adult piscivores\\
        $s$ & $0.5$ & survival rate of adult and juvenile piscivores over the winter period\\
    \bottomrule
    \end{tabular}
\end{table*}
Depending on the harvest rate the system settles into a piscivore- or planktivore-dominated state. In the first mentioned scenario the planktivore abundance is kept at a low level because of a large occurrence of adult piscivores whereas in the second scenario the large population of planktivores hinders the piscivore population to grow because the planktivores' predation of the juvenile group.\\
We focus on regime shifts from the piscivore- into the planktivore-dominated state due to increasing harvest rate or angling pressure $qE$. In figure \ref{fig: state space} the key features of the dynamics are illustrated in state space for a more detailed mathematical description.\\
\section{Numerical methods}\label{sec: method}
In subsection \ref{subsec: BL estimation} and \ref{subsec: statistical measures} we introduce the Bayesian drift slope estimation procedure as well as the methods used to estimate the statistical measures. Finally, in subsection \ref{subsec: model comparison} we describe our approach of trend-significance testing via Bayes factors.\\
\subsection{Drift slope estimation scheme}\label{subsec: BL estimation}
Starting with the Langevin equation \ref{eq:langevin} we parameterize the drift and diffusion as $h(x(t), t) \equiv h(x(t))$ and  $g(x(t), t) \equiv const. =: \sigma$.
Since we assume to be in a fixed point and close to a bifurcation we develop $h(x,t)$ into a Taylor series up to order three which is sufficient to describe the normal forms of simple bifurcation scenarios (\cite{b:strogatz}). This results in
\begin{align}
\begin{split}
h(x(t),t) &= \alpha_0(t) + \alpha_1(t) (x - x^*) + \alpha_2(t) (x - x^*)^2 \\
&+ \alpha_3(t) (x - x^*)^3 + \mathcal{O}((x-x^*)^4)  \label{eq:taylor} 
\end{split},
\end{align}
so that the information on the linear stability is incorporated in $\alpha_1$. For practical reasons equation \ref{eq:taylor} is used in the form
\begin{align}
\begin{split}
    h_{\rm MC}(x(t),t) &= \theta_0 (t; x^*) + \theta_1 (t; x^*) \cdot x + \theta_2 (t; x^*) \cdot x^2 \\ 
    &+ \theta_3 (t; x^*) \cdot x^3 + \mathcal{O}(x^4)
    \end{split}
\end{align}
in the numerical approach, where an arbitrary fixed point $x^*$ is incorporated in the coefficients $\underline{\theta}$ by algebraic transformation and comparison of coefficients. A change of the negative sign of the slope 
\begin{linenomath*}
\begin{equation}
\zeta = \left.\frac{\text{d}h(x)}{\text{d}x}\right\vert_{x = x^*} 
\end{equation}
\end{linenomath*}
of the nonlinear drift at the fixed point $x^*$ which is estimated to be the data mean corresponds to a loss of stability via the formalism of linear stability analysis (\cite{a:Hessler2021}).\\
The task is now to estimate the parameters $\underline{\theta}$. Their posterior distribution is given by applying Bayes' theorem
\begin{linenomath*}
\begin{equation}
p(\underline{\theta},\sigma |\underline{d}, \mathcal{I}) = \frac{p(\underline{d}|\underline{\theta}, \sigma, \mathcal{I}) \cdot p(\underline{\theta}, \sigma |\mathcal{I})}{p(\underline{d}  | \mathcal{I})} .
\end{equation}
\end{linenomath*}
The likelihood $p(\underline{d}|\underline{\theta}, \sigma, \mathcal{I})$ is given as the transition probability of the process defined by equation \eqref{eq:langevin} (see \cite{a:Hessler2021}) and the  prior knowledge is incorporated in $p(\underline{\theta}, \sigma |\mathcal{I})$. The evidence $p(\underline{d}  | \mathcal{I})$ normalizes the posterior probability density function (pdf) $p(\underline{\theta},\sigma |\underline{d}, \mathcal{I})$. One advantage of this procedure is the consistent definition of credibility bands of the estimated parameters based on the posterior pdf. The posterior distribution of the  parameters can be estimated via MCMC sampling with the flat Jeffreys' priors 
\begin{align}
p_{\rm prior}(\theta_0,\theta_1) = \frac{1}{2\pi (1+\theta_1^2)^\frac{3}{2}}
\end{align}
and
\begin{align}
p_{\rm prior}(\sigma ) = \frac{1}{\sigma}
\end{align}
for the scale variable $\sigma$ (\cite{b:linden2014}). Gaussian priors \begin{equation}
\begin{aligned}
p_{\rm prior}(\theta_2) &= \mathcal{N}(\mu, \sigma_{\theta_2}),  \\
p_ {\rm prior}(\theta_3) &= \mathcal{N}(\mu, \sigma_{\theta_3})
\end{aligned}
\end{equation}
centred around the mean $\mu = 0$ with standard deviations $\sigma_{\theta_i}$ in an adequate range are used for the rest of the parameters. The flat Jeffreys' priors are chosen broadly as $[-50, 50]$ for $\theta_{0,1}$ and $[0,50]$ for $\theta_4$ except for the analysis of the deseasonalized versions of the correlated models. In these cases (red lines in D-I) the prior range is chosen even broader as $[-70, 70]$ for $\theta_{0,1}$ and $[0,70]$ for $\theta_4$ to make sure that the available data determines the posterior distribution. The Gaussian priors for $\theta_{2,3}$ are implemented with $\sigma = \lbrace 4,\ 8 \rbrace$, respectively.\\ 
We use the MCMC sampling algorithm implemented in the python package \textit{emcee} (\cite{a:Foreman-Mackey2013}). The method is applied in rolling windows in order to resolve the time evolution of the drift slope. A detailed description of the presented algorithm and its implementation steps can be found in \cite{a:Hessler2021}.\\
\subsection{Statistical leading indicators}\label{subsec: statistical measures}
The biased autocorrelation at lag-1 is computed via \textit{statsmodels.tsa.stattools.acf} (\cite{ipr:seabold2010}) and the biased standard deviation $\hat{\sigma}$ via \textit{numpy.std} (\cite{a:Harris2020}). The skewness $\gamma$ and kurtosis $\omega$ calculations are performed with the biased uncorrected estimators of the python package \textit{scipy.stats} (\cite{a:scipy2020}). The biased versions are used, because of the large sample sizes which provide sufficient accuracy. The skewness definition follows the not-adjusted Fisher-Pearson estimator and the kurtosis is defined via the Pearson estimator corresponding to a kurtosis $\omega = 3$ for a Gaussian distribution.\\
\subsection{Significance testing via Bayesian model comparison}\label{subsec: model comparison}
A Bayesian model comparison is used in order to quantify the significance of the various leading indicators. Shortly summarized, we compare the probability that the estimated leading indicators can be explained by a linear trend model to the probability that the measures are described by a constant model via the concept of Bayes factors (\cite{b:Jeffreys1998}). The Bayes factors
\begin{equation}
\begin{aligned}
    BF_{ij} = \frac{p(\underline{\mathcal{I}} | \mathcal{M}_{i})}{p(\underline{\mathcal{I}} | \mathcal{M}_{j})} \textrm{ with }i,j \in \lbrace 1, 2\rbrace\textrm{ and }i \neq j
    \end{aligned}
\end{equation} 
 are computed in the same way for the different leading indicator candidates. The prior parameter ranges of the linear model $\mathcal{M}_1$ and the constant model $\mathcal{M}_2$ are adapted to the specific leading indicator time series data based on the following procedure:
\begin{enumerate}
    \item The initial value $\mathcal{I}_0$ of each leading indicator time series is used as mean $\mu_{\mathcal{M}_2}$ of a Gaussian distribution $\mathcal{N}(\mu_{\mathcal{M}_2}=\mathcal{I}_0, \sigma_{\mathcal{M}_2} = 1)$. The distribution is used to draw the parameter of the constant model $\mathcal{M}_2$ which are also used as the intercepts $b$ of the linear model $\mathcal{M}_1: y = a \cdot x + b$.
    \item The slope $a$ of the linear model $\mathcal{M}_1$ is drawn from a uniform distribution in the range $[0, 1.5 \cdot\frac{\max(\underline{\mathcal{I}})-\min(\underline{\mathcal{I}})}{t_{\rm end}-t_{\rm start}}]$. The calculations of BFs in the case of deceasing skewness is performed by drawing uniformly from the interval $[- 1.5 \cdot\frac{\max(\underline{\mathcal{I}})-\min(\underline{\mathcal{I}})}{t_{\rm end}-t_{\rm start}}, 0]$.
    \item The logarithmic noise is drawn uniformly from the interval $[\log(0.5), \log(5)]$.
\end{enumerate}
The convergence of the results is guaranteed by drawing $10^7$ realisations of each model in each BF calculation.
\section{Results}\label{sec: results}
\subsection{Drift slope analysis}\label{subsec: drift slope analysis}
\begin{figure*}[!htbp]
\includegraphics[width=0.9 \textwidth]{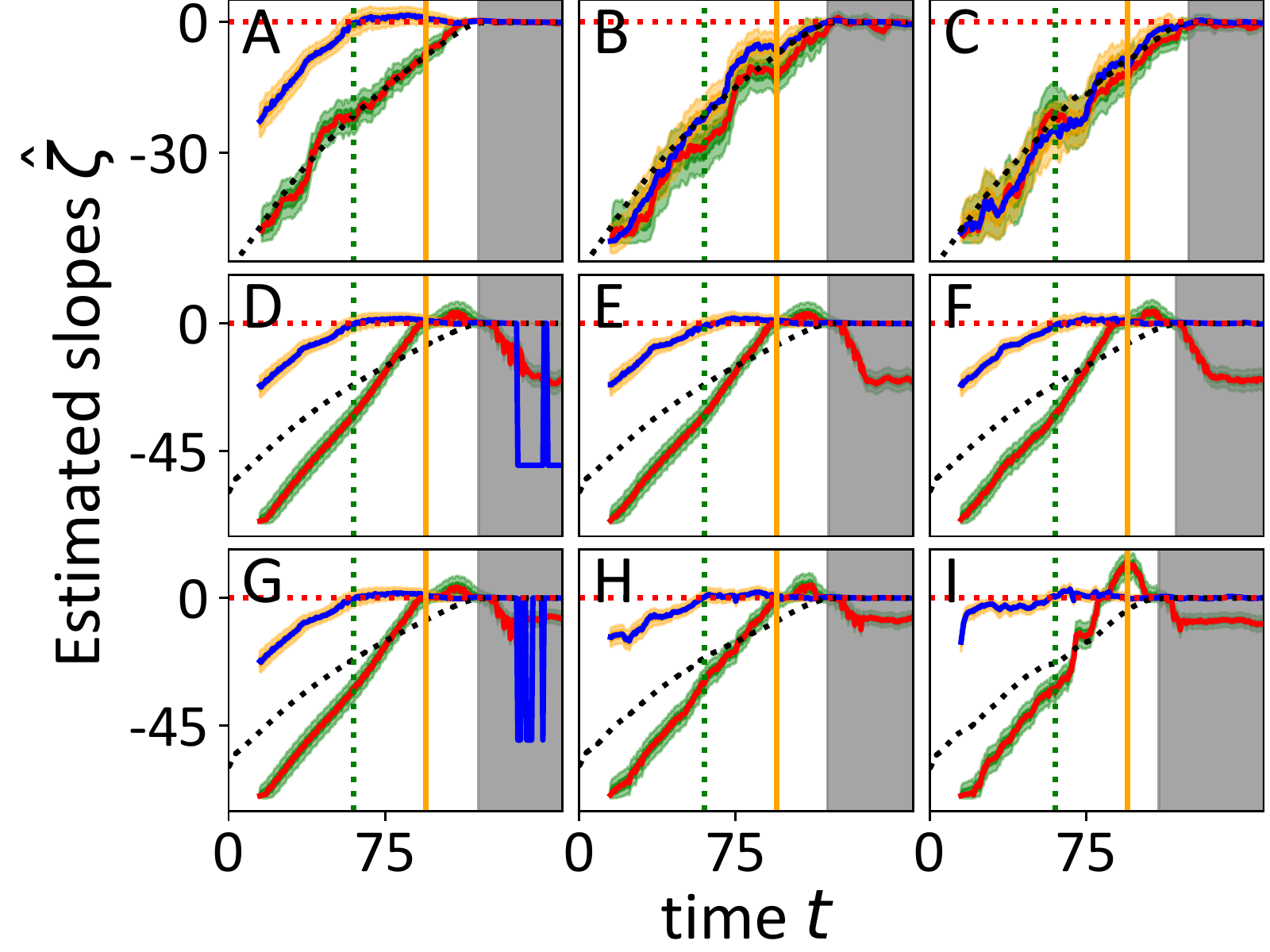}
\caption[Drift slope analysis]{Results of the drift slope analysis for the ecological model with white (A-C), pink (D-F) and red noise (G-I). The columns from left to right correspond to the noise levels $\sigma = \lbrace 0.1,\ 2.2,\ 4.5\rbrace$. The computations are performed on the time series without deseasonalization (blue lines with orange credibility bands) and with preparation by deseasonalizing the data (red lines with green credibility bands). The green dotted and the orange solid vertical lines indicate the attractor switch point of the deterministic system and the point of no return, respectively, that is defined as the time at which the destabilization cannot be stopped by reducing the harvest rate to $qE = 0.1$. A comparison of the drift slope estimates $\hat{\zeta}$ to the analytical partial planktivore $F$-drift slopes (black dotted lines) confirms that the estimates are quantivatively accurate in the case of white noise and qualitatively reasonable for correlated noise. The deseasonalized versions exhibit clear trends and reach the critical zero marked by the red dotted horizontal line around the transition time that is approximately signed by the beginning of the grey shaded area that is defined as the time at which the small planktivore population counts more than $21$ individuals. Although, the method is not designed to deal with correlated noise and non-Markovian time series the seasonality of the data has much more influence than the correlated noise. The seasonality reduces clearness of the trends and leads to an earlier zero crossing of the drift slope for weak white and all correlated noise scenarios. In the weak white noise case the numerical method seems to interpret the seasonal effects incorrectly, probably as noise influence. For bigger noise levels the seasonal effects become insignificant for the white noise cases, but not for the correlated noise scenarios. The strong fluctuation of the drift slope estimates in the post transition region of the subfigures (D, G) are probably due to the small correlated noise contributions in the new stable state.}
\label{fig: drift slope results}
\end{figure*}

The drift slope estimation method that is shortly summarized in section \ref{sec: method} and described in detail in \cite{a:Hessler2021} is applied to time series simulations of the seasonal ecological model with white, pink and red noise each of which is realised for three noise levels $\sigma = \lbrace 0.1,\ 2.2,\ 4.5\rbrace$. The data is evaluated in windows of $750$ data points that are shifted by $30$ points per step and analysed in two szenarios: First, without pre-processing of the data by deseasonalization and second, with a deseasonalization before computing the drift slope. The analysis results are presented in figure \ref{fig: drift slope results}. The results of the first approach are marked in blue with orange credibility bands defined as the $\SI{16}{\percent}$ to $\SI{84}{\percent}$ and $\SI{1}{\percent}$ to $\SI{99}{\percent}$ percentile of the drift slope posterior modelled by a kernel density estimate of the sampled parameters (\cite{a:sklearn}). The second ansatz is shown in red with the corresponding green credibility bands. Both estimation results can be compared to the analytical values of the smoothed partial derivative of the planktivore $F$-drift in planktivore $F$-direction shown as a black dotted line and computed with the data of the model realisations (cf. supplementary material (\cite{a:suppleco})). The green dotted and orange solid vertical lines are defined equivalently to \cite{a:Perretti2012} as the attractor switch point and the ``point of no return'', respectively, which is defined as the year in which even a reduction of the harvest rate to $qE = 0.1$ does not inhibit the destabilization process of the ecological system. The beginning of the grey shaded area is a subjectively defined time at which the previously small planktivore population exceeds $21$ individuals and serves as an orientation for the ongoing destabilization process. Each column from left to right belongs to one of the three noise levels $\sigma = \lbrace 0.1,\ 2.2,\ 4.5\rbrace$. The first (A-C), second (D-F) and third row (G-I) contain the drift slope results of the realisations of the model with additional white, pink and red noise, respectively. By comparing the results of the analyses with and without deseasonalization over various noise environments of the model we gain valuable insights into the capacities an limits of the methodological concept: In the figures \ref{fig: drift slope results} (A-C) the deseasonalized cases perform rather similar to the cases without deseasonalization apart from the weak noise case (A) with $\sigma = 0.1$. This leads to the conclusion that in the weak noise case (A) the seasonal effects in the data are interpreted by the model probably in terms of noise fluctuations because the parameterization cannot capture the predominant seasonality. With increasing noise the seasonal effects become insignificant as visible in the figures \ref{fig: drift slope results} (B, C) because the noise level covers and hides the seasonal component of the data. The drift slope indicator seems to be suitable to provide information about the resilience of this ecological model with white noise, whereas seasonal aspects should be treated carefully for small noise levels. A  comparison with the analytical partial derivative of the planktivore $F$-drift reveals that the deseasonalized estimates are accurate in the white noise cases. 
Technically, the drift slope estimation is not designed in order to deal with correlated noise and thus, with Non-Markovianity. As expected by that fact, the drift slope estimates in (D-I) exhibit a systematic quantitative estimation error compared to the analytical partial planktivore $F$-derivative. Nevertheless, the qualitative trends remain unchanged. Since the trend resolution towards a potentially zero-crossing is the crucial feature in terms of leading indicator use of the drift slope $\zeta$ the findings in (D-I) show that it can be still a helpful early warning tool even in highly correlated and noisy situations. Similar to the results in the weak white noise case (A) the results reach the critical zero line around the attractor switch point and exhibit less clear trends as their deseasonalized counterparts that reach the critical zero around the actual transition that is approximately marked by the beginning of the grey shaded area. In contrast to the white noise cases the seasonality in the correlated noise cases influences the results for all noise levels in a similar way: Without deseasonalization the drift slope reaches zero around the attractor switch point whereas it reaches zero around the point of no return in the absence of seasonality. The critical zero crossing of the drift slope in the deseasoned versions of the correlated noise cases seems to be a bit earlier than the crossings of the white noise counterparts. The high impact of the seasonality in the strong correlated noise cases compared to the strong white noise cases (B, C) is due to the correlation of the noise itself: The noise correlation tends to amplify or weaken the annual amplitudes, whereby the seasonal component of the time series is not hidden by the noise, but more or less preserved. Note also, that there is no clear formal reason for the zero crossing of the blue drift slopes at the attractor switch point or for the red drift slopes reaching the critical zero around the point of no return in the correlated noise cases. Besides, the strong fluctuating slope estimates after the transition time in figure \ref{fig: drift slope results} (D, G) without deseasonalization are numerical artefacts probably caused by the small correlated noise contributions in the new stable state.\\
In conclusion, the drift slope trends are rather robust in the presented model cases and provide reliable information about the resilience and destabilization of the ecological system. The method is relatively complicated to implement in contrast to leading indicator candidates as the AR1 or the std $\hat{\sigma}$. Anyhow, its performance and robustness could be important advantages in the field of ecology and other data-driven research as outlined in the next subsection \ref{subsec: comparison} in which the performance of the drift slope in this dynamical rolling window setting is compared to common leading indicator candidates.
\subsection{Comparison of leading indicators' performance} \label{subsec: comparison}
In order to compare the performance of the drift slope indicator with established early warning candidates as the autocorrelation at lag-1 (AR1), the standard deviation (std) $\hat{\sigma}$, the skewness $\gamma$ or the kurtosis $\omega$ we use a Bayesian model comparison in which we compute the Bayes factors $BF_{ij}$ with $i,j \in \lbrace 1, 2\rbrace$ and $i \neq j$ that are defined as the ratio 
\begin{equation}
\begin{aligned}
    BF_{ij} = \frac{p(\underline{\mathcal{I}} | \mathcal{M}_{i})}{p(\underline{\mathcal{I}} | \mathcal{M}_{j})}
    \end{aligned}
\end{equation} 
of the evidences $p(\underline{\mathcal{I}} | \mathcal{M}_{1,2})$ that a linear trend model (model $\mathcal{M}_1$) or a constant model (model $\mathcal{M}_2$) explain the leading indicator datasets $\underline{\mathcal{I}}$ up to the ``point of no return''. The $BF_{ij}$ are calculated for each of the above mentioned noise levels, noise types and the datasets without and with deseasonalization. A Bayes factor is declared to be significant for $BF_{ij} > 100$ (\cite{b:Jeffreys1998}) to take into account the fact that most of the Bayes factors lie in the range $10 < BF_{ij} < 100$ or are significantly bigger than $100$. The results of the comparison without deseasonalizing the data are summarized in table \ref{table: performance comparison without deseasonalization} where the color code follows \cite{a:Hessler2021} with a significant $BF_{12}$ or $BF_{21}$ marked by green and orange tiles, respectively, and grey tiles denote cases in which none of the models is favourable. The results of the kurtosis $\omega$ are excluded from further discussion, because of the ambiguous, non-monotone and very noisy trends with jumps which cannot be reliably interpreted by eye or captured by the linear model $\mathcal{M}_1$ of the Bayes model comparison. In some cases the constant model was erroneously preferred or the results were not significant. The corresponding curves of the leading indicators of each case can be found in the supplementary material (\cite{a:suppleco}). Bayes factor pairs with infinite and zero entries correspond to one of the two models with evidence of zero and thus, the model with finite evidence is preferred.  Without deseasonalization the common leading indicators AR1, std $\hat{\sigma}$ and skewness $\gamma$ do not exhibit a significant slope following the Bayesian model comparison in most of the cases, although the time series resolution is relatively high (\cite{a:Perretti2012}) and the time windows are chosen as big as in the last subsection \ref{subsec: drift slope analysis}. Without deseasonalization the AR1 just performs well in the white noise cases with $\sigma = \lbrace 2.2, 4.5 \rbrace$, whereas the skewness $\gamma$ does not exhibit any reliable pattern of applicability. Note, that these results remain unchanged if the data is only detrended, but not deseasonalized. The corresponding analysis can be found in the supplementary material (\cite{a:suppleco}). If the results are compared to the deseasonalized counterparts of table \ref{table: performance comparison with deseasonalization} the green tile of std $\hat{\sigma}$ and the significant white noise cases of the skewness $\gamma$ turn out to be artefacts caused by the seasonal nature of the time series. Interestingly, the deseasonalization leads to a consistent significance pattern of the skewness $\gamma$ if only the correlated noise cases are considered. Therefore, the general applicability of the skewness $\gamma$ as leading indicator is ill-advised since it is rather sensitive to noise types, seasonality and e.g. bistability of the system. Nevertheless, it could be useful under specific conditions as the correlated noise cases considered here or in flickering regimes of bistable systems. Only the recently proposed drift slope and the AR1 with deseasonalization seem to yield reliable results. The performance of the AR1 is significantly improved by deseasonalization that leads to significant trends in all cases as suggested by a comparison of the tables \ref{table: performance comparison with deseasonalization} and \ref{table: performance comparison without deseasonalization}. Under the same conditions the drift slope turns out to be not very sensitive to the seasonal character of the data apart from the early plateaus discussed in subsection \ref{subsec: drift slope analysis}. The drift slope $\zeta$ leads to significant positive trends in all considered cases without distinction of non-deseasonalized and deseasonalized data. The results confirm in most instances the results of \cite{a:Perretti2012} where a very poor applicability of the standard leading indicator candidates to the ecological test dataset is observed. The most robust leading indicator under strong noise was found to be the variance or std $\hat{\sigma}$ in \cite{a:Perretti2012}. The Bayes factor analysis proposes AR1 to be the most reliable indicator of the standard measures and rejects the std $\hat{\sigma}$ as a robust indicator.\\
Following the results of this study the drift slope $\zeta$ is a possible leading indicator candidate also in very noisy situations, provided that a suitable sampling rate of the time series is guaranteed. In the next subsection \ref{subsec: ws limits} the limitations of the drift slope estimates $\hat{\zeta}$ and their sensitivity to small window sizes are investigated because, as stated in \cite{a:Perretti2012}, ecological time series are often short and possible window sizes are strongly limited by that fact.\\
\begin{landscape}
\begin{table*}
\ra{1.4}
   \centering
        \caption{Summary of the Bayes factors comparing a linear model $\mathcal{M}_1$ with positive slope to a constant model $\mathcal{M}_2$ for the drift slope $\zeta$, the AR1, the std $\hat{\sigma}$ and the skewness $\gamma$ for various noise types and levels without deseasonalization of the data. The kurtosis $\omega$ is excluded because of non-monotone behaviour. Green tiles mark a $BF_{12} > 100$ (\cite{b:Jeffreys1998}) which is the threshold for a significant leading indicator trend. Grey tiles mark insignificant results. The constant model $\mathcal{M}_2$ is never preferred in the analysis. Infinite Bayes factors result from one model with evidence zero which leads to preferring the finite evidence model. Only the drift slope $\zeta$ performs well in the considered cases. The AR1 has a very limited applicability and the skewness $\gamma$ is not reliable over all cases. The green tile of the std $\hat{\sigma}$ is an artefact of the seasonality which is confirmed by a comparison with table \ref{table: performance comparison with deseasonalization}. For completeness, the same analysis is performed for the skewness with a linear model $\mathcal{M}_1$ with negative slope in the supplementary material (\cite{a:suppleco}).}
    \label{table: performance comparison without deseasonalization}
    \begin{NiceTabular}{@{}lrrrrrrrrr}[]\toprule[1.3pt]
     & \multicolumn{3}{c}{white noise ($P_{\rm white} \approx 7.5$)} & \multicolumn{3}{c}{pink noise ($P_{\rm pink} \approx 2 \cdot P_{\rm white}$)} & \multicolumn{3}{c}{red noise ($P_{\rm red} \approx P_{\rm pink}$)} \\
     \cmidrule(lr){2-4} \cmidrule(lr){5-7} \cmidrule(lr){8-10}
    noise level &  $\sigma = 0.1$ & $\sigma = 2.2$ & $\sigma = 4.5$ &  $\sigma = 0.1$ & $\sigma = 2.2$ & $\sigma = 4.5$ &  $\sigma = 0.1$ & $\sigma = 2.2$ & $\sigma = 4.5$ \\ 
    \cmidrule{1 - 10} 
    indicator $\underline{\mathcal{I}}$
    & \begin{tabular}{@{}r@{}} $\rm BF_{12}$ \\ $\rm BF_{21}$ \end{tabular} 
    & \begin{tabular}{@{}r@{}} $\rm BF_{12}$ \\ $\rm BF_{21}$ \end{tabular} 
    & \begin{tabular}{@{}r@{}} $\rm BF_{12}$ \\ $\rm BF_{21}$ \end{tabular}
    & \begin{tabular}{@{}r@{}} $\rm BF_{12}$ \\ $\rm BF_{21}$ \end{tabular} 
    & \begin{tabular}{@{}r@{}} $\rm BF_{12}$ \\ $\rm BF_{21}$ \end{tabular} 
    & \begin{tabular}{@{}r@{}} $\rm BF_{12}$ \\ $\rm BF_{21}$ \end{tabular} 
    & \begin{tabular}{@{}r@{}} $\rm BF_{12}$ \\ $\rm BF_{21}$ \end{tabular}
    & \begin{tabular}{@{}r@{}} $\rm BF_{12}$ \\ $\rm BF_{21}$ \end{tabular}
    & \begin{tabular}{@{}r@{}} $\rm BF_{12}$ \\ $\rm BF_{21}$ \end{tabular}
     \\ 
    \midrule
    slope $\zeta$
    & \cellcolor{green}\begin{tabular}{@{}r@{}} $\infty$ \\ 0 \end{tabular}
    & \cellcolor{green}\begin{tabular}{@{}r@{}} $\infty$ \\ 0 \end{tabular}
    & \cellcolor{green}\begin{tabular}{@{}r@{}} $\infty$ \\ 0 \end{tabular}
    & \cellcolor{green}\begin{tabular}{@{}r@{}} $\infty$ \\ 0 \end{tabular}
    & \cellcolor{green}\begin{tabular}{@{}r@{}} $\infty$ \\ 0 \end{tabular}
    & \cellcolor{green}\begin{tabular}{@{}r@{}} $\infty$ \\ 0 \end{tabular}
    & \cellcolor{green}\begin{tabular}{@{}r@{}} $\infty$ \\ 0 \end{tabular}
    & \cellcolor{green}\begin{tabular}{@{}r@{}} $3.3 \cdot 10^{99}$ \\ $3.0 \cdot 10^{-100}$ \end{tabular}
    & \cellcolor{green}\begin{tabular}{@{}r@{}} $3.0 \cdot 10^{92}$ \\ $3.3 \cdot 10^{-93}$ \end{tabular} \\
    AR1
    & \cellcolor{gray}\begin{tabular}{@{}r@{}} $12$ \\ $8.3\cdot 10^{-2}$ \end{tabular}
    & \cellcolor{green}\begin{tabular}{@{}r@{}} $8.8 \cdot 10^{7}$ \\ $1.1 \cdot 10^{-8}$ \end{tabular}
    & \cellcolor{green}\begin{tabular}{@{}r@{}} $3.8 \cdot 10^{7}$ \\ $2.6 \cdot 10^{-8}$ \end{tabular}
    & \cellcolor{gray}\begin{tabular}{@{}r@{}} $9.4$ \\ $0.11$ \end{tabular}
    & \cellcolor{gray}\begin{tabular}{@{}r@{}} $8.7$ \\ $0.11$ \end{tabular}
    & \cellcolor{gray}\begin{tabular}{@{}r@{}} $9.1$ \\ $0.11$ \end{tabular}
    & \cellcolor{gray}\begin{tabular}{@{}r@{}} $9.5$ \\ $0.11$ \end{tabular}
    & \cellcolor{gray}\begin{tabular}{@{}r@{}} $3.1$ \\ $0.32$ \end{tabular}
    & \cellcolor{gray}\begin{tabular}{@{}r@{}} $1.2$ \\ $0.85$ \end{tabular} \\
    std $\hat{\sigma}$
    & \cellcolor{gray}\begin{tabular}{@{}r@{}} $19$ \\ $5.3\cdot 10^{-2}$ \end{tabular}
    & \cellcolor{gray}\begin{tabular}{@{}r@{}} $18$ \\ $5.5 \cdot 10^{-2}$ \end{tabular}
    & \cellcolor{gray}\begin{tabular}{@{}r@{}} $57$ \\ $1.8 \cdot 10^{-2}$ \end{tabular}
    & \cellcolor{gray}\begin{tabular}{@{}r@{}} $18$ \\ $5.5 \cdot 10^{-2}$ \end{tabular}
    & \cellcolor{gray}\begin{tabular}{@{}r@{}} $18$ \\ $5.5\cdot 10^{-2}$ \end{tabular}
    & \cellcolor{gray}\begin{tabular}{@{}r@{}} $39$ \\ $2.6 \cdot 10^{-2}$ \end{tabular}
    & \cellcolor{gray}\begin{tabular}{@{}r@{}} $19$ \\ $5.4 \cdot 10^{-2}$ \end{tabular}
    & \cellcolor{gray}\begin{tabular}{@{}r@{}} $10$ \\ $0.10$ \end{tabular}
    & \cellcolor{green}\begin{tabular}{@{}r@{}} $8.3 \cdot 10^{4}$ \\ $1.2 \cdot 10^{-5}$ \end{tabular} \\
    skewness $\gamma$
    & \cellcolor{green}\begin{tabular}{@{}r@{}} $6.1\cdot 10^{6}$ \\ $1.6\cdot 10^{-7}$ \end{tabular}
    & \cellcolor{green}\begin{tabular}{@{}r@{}} $1.3 \cdot 10^6$ \\ $7.9\cdot 10^{-7}$ \end{tabular}
    & \cellcolor{gray}\begin{tabular}{@{}r@{}} $0.91$ \\ $1.1$ \end{tabular}
    & \cellcolor{green}\begin{tabular}{@{}r@{}} $1.1\cdot 10^7$ \\ $9.1\cdot 10^{-8}$ \end{tabular}
    & \cellcolor{green}\begin{tabular}{@{}r@{}} $9.5\cdot 10^5$ \\ $1.1\cdot 10^{-6}$ \end{tabular}
    & \cellcolor{green}\begin{tabular}{@{}r@{}} $8.3 \cdot 10^{8}$ \\ $1.2 \cdot 10^{-9}$ \end{tabular}
    & \cellcolor{green}\begin{tabular}{@{}r@{}} $1.1\cdot 10^7$ \\ $8.8\cdot 10^{-8}$ \end{tabular}
    & \cellcolor{gray}\begin{tabular}{@{}r@{}} $61$ \\ $1.6\cdot 10^{-2}$ \end{tabular}
& \cellcolor{green}\begin{tabular}{@{}r@{}} $3.4 \cdot 10^{4}$ \\ $3.0 \cdot 10^{-5}$ \end{tabular} \\
\bottomrule
\end{NiceTabular}
\end{table*}
\end{landscape}
\vspace{25pt}
\begin{landscape}
\begin{table*}
\caption{Same as table \ref{table: performance comparison without deseasonalization} with deseasonalization of the data. The drift slope $\zeta$ applies as before to all test cases. However, deseasonalization improves the performance of the AR1 as leading indicator significantly as it also works for all deseasonalized cases, whereas the std $\hat{\sigma}$ does not work. This leads to the conclusion that the fragmentary applicability of the std $\hat{\sigma}$ in table \ref{table: performance comparison without deseasonalization} is an artefact due to a misinterpretation of the seasonal character of the time series. Confronted with these results it is important to keep in mind that the AR1 is only a reliable indicator for a bifurcation-induced critical transition if the std increases at the same time. We could not find one case in which the Bayes factors of both AR1 and std are decisive at the same time. The skewness becomes a reliable indicator for the considered correlated noise cases, whereas its positive trends disappear in the white noise cases due to the deseasonalization procedure. The same analysis for the skewness with a linear model $\mathcal{M}_1$ with negative slope can be found in the supplementary material (\cite{a:suppleco}).}
\label{table: performance comparison with deseasonalization}
\ra{1.4}
  \centering
    \begin{NiceTabular}{@{}lrrrrrrrrr}[]\toprule[1.3pt]
     & \multicolumn{3}{c}{white noise ($P_{\rm white} \approx 7.5$)} & \multicolumn{3}{c}{pink noise ($P_{\rm pink} \approx 2 \cdot P_{\rm white}$)} & \multicolumn{3}{c}{red noise ($P_{\rm red} \approx P_{\rm pink}$)} \\
     \cmidrule(lr){2-4} \cmidrule(lr){5-7} \cmidrule(lr){8-10}
    noise level &  $\sigma = 0.1$ & $\sigma = 2.2$ & $\sigma = 4.5$ &  $\sigma = 0.1$ & $\sigma = 2.2$ & $\sigma = 4.5$ &  $\sigma = 0.1$ & $\sigma = 2.2$ & $\sigma = 4.5$ \\ 
    \cmidrule{1 - 10} 
    indicator $\underline{\mathcal{I}}$
    & \begin{tabular}{@{}r@{}} $\rm BF_{12}$ \\ $\rm BF_{21}$ \end{tabular} 
    & \begin{tabular}{@{}r@{}} $\rm BF_{12}$ \\ $\rm BF_{21}$ \end{tabular} 
    & \begin{tabular}{@{}r@{}} $\rm BF_{12}$ \\ $\rm BF_{21}$ \end{tabular}
    & \begin{tabular}{@{}r@{}} $\rm BF_{12}$ \\ $\rm BF_{21}$ \end{tabular} 
    & \begin{tabular}{@{}r@{}} $\rm BF_{12}$ \\ $\rm BF_{21}$ \end{tabular} 
    & \begin{tabular}{@{}r@{}} $\rm BF_{12}$ \\ $\rm BF_{21}$ \end{tabular} 
    & \begin{tabular}{@{}r@{}} $\rm BF_{12}$ \\ $\rm BF_{21}$ \end{tabular}
    & \begin{tabular}{@{}r@{}} $\rm BF_{12}$ \\ $\rm BF_{21}$ \end{tabular}
    & \begin{tabular}{@{}r@{}} $\rm BF_{12}$ \\ $\rm BF_{21}$ \end{tabular}
     \\ 
    \midrule
    slope $\zeta$
    & \cellcolor{green}\begin{tabular}{@{}r@{}} $\infty$ \\ 0 \end{tabular}
    & \cellcolor{green}\begin{tabular}{@{}r@{}} $\infty$ \\ 0 \end{tabular}
    & \cellcolor{green}\begin{tabular}{@{}r@{}} $\infty$ \\ 0 \end{tabular}
    & \cellcolor{green}\begin{tabular}{@{}r@{}} $\infty$ \\ 0 \end{tabular}
    & \cellcolor{green}\begin{tabular}{@{}r@{}} $\infty$ \\ 0 \end{tabular}
    & \cellcolor{green}\begin{tabular}{@{}r@{}} $\infty$ \\ 0 \end{tabular}
    & \cellcolor{green}\begin{tabular}{@{}r@{}} $\infty$ \\ 0 \end{tabular}
    & \cellcolor{green}\begin{tabular}{@{}r@{}} $\infty$ \\ 0 \end{tabular}
    & \cellcolor{green}\begin{tabular}{@{}r@{}} $\infty$ \\ 0 \end{tabular}\\
    AR1
    & \cellcolor{green}\begin{tabular}{@{}r@{}} $3.4\cdot 10^{5}$ \\ $3.0 \cdot 10^{-6}$ \end{tabular}
    & \cellcolor{green}\begin{tabular}{@{}r@{}} $1.6 \cdot 10^{7}$ \\ $6.2 \cdot 10^{-8}$ \end{tabular}
    & \cellcolor{green}\begin{tabular}{@{}r@{}} $6.1 \cdot 10^{6}$ \\ $1.6 \cdot 10^{-7}$ \end{tabular}
    & \cellcolor{green}\begin{tabular}{@{}r@{}} $4.3\cdot 10^{8}$ \\ $2.3\cdot 10^{-9}$ \end{tabular}
    & \cellcolor{green}\begin{tabular}{@{}r@{}} $1.5\cdot 10^{8}$ \\ $6.6\cdot 10^{-9}$ \end{tabular}
    & \cellcolor{green}\begin{tabular}{@{}r@{}} $1.4\cdot 10^{9}$ \\ $7.3\cdot 10^{-10}$ \end{tabular}
    & \cellcolor{green}\begin{tabular}{@{}r@{}} $4.3\cdot 10^{8}$ \\ $2.3\cdot 10^{-9}$ \end{tabular}
    & \cellcolor{green}\begin{tabular}{@{}r@{}} $1.4\cdot 10^{8}$ \\ $7.2\cdot 10^{-9}$ \end{tabular}
    & \cellcolor{green}\begin{tabular}{@{}r@{}} $1.2\cdot 10^{9}$ \\ $8.6\cdot 10^{-10}$ \end{tabular} \\
    std $\hat{\sigma}$
    & \cellcolor{gray}\begin{tabular}{@{}r@{}} $1.0$ \\ $1.0$ \end{tabular}
    & \cellcolor{gray}\begin{tabular}{@{}r@{}} $5.8$ \\ $0.17$ \end{tabular}
    & \cellcolor{gray}\begin{tabular}{@{}r@{}} $40$ \\ $2.5 \cdot 10^{-2}$ \end{tabular}
    & \cellcolor{gray}\begin{tabular}{@{}r@{}} $1.1$ \\ $0.89$ \end{tabular}
    & \cellcolor{gray}\begin{tabular}{@{}r@{}} $1.1$ \\ $0.89$ \end{tabular}
    & \cellcolor{gray}\begin{tabular}{@{}r@{}} $1.1$ \\ $0.88$ \end{tabular}
    & \cellcolor{gray}\begin{tabular}{@{}r@{}} $1.1$ \\ $0.89$ \end{tabular}
    & \cellcolor{gray}\begin{tabular}{@{}r@{}} $1.1$ \\ $0.92$ \end{tabular}
    & \cellcolor{gray}\begin{tabular}{@{}r@{}} $1.3$ \\ $0.76$ \end{tabular} \\
    skewness $\gamma$
    & \cellcolor{gray}\begin{tabular}{@{}r@{}} $14$ \\ $7.1\cdot 10^{-2}$ \end{tabular}
    & \cellcolor{gray}\begin{tabular}{@{}r@{}} $0.58$ \\ $1.7$ \end{tabular}
    & \cellcolor{gray}\begin{tabular}{@{}r@{}} $0.44$ \\ $2.8$ \end{tabular}
    & \cellcolor{green}\begin{tabular}{@{}r@{}} $1.6\cdot 10^{90}$ \\ $6.3\cdot 10^{-91}$ \end{tabular}
    & \cellcolor{green}\begin{tabular}{@{}r@{}} $3.6\cdot 10^{89}$ \\ $2.8\cdot 10^{-90}$ \end{tabular}
    & \cellcolor{green}\begin{tabular}{@{}r@{}} $1.1\cdot 10^{89}$ \\ $8.8\cdot 10^{-90}$ \end{tabular}
    & \cellcolor{green}\begin{tabular}{@{}r@{}} $1.8\cdot 10^{90}$ \\ $5.7\cdot 10^{-91}$ \end{tabular}
    & \cellcolor{green}\begin{tabular}{@{}r@{}} $2.1\cdot 10^{89}$ \\ $4.7\cdot 10^{-90}$ \end{tabular}
    & \cellcolor{green}\begin{tabular}{@{}r@{}} $2.3\cdot 10^{93}$ \\ $4.4\cdot 10^{-94}$ \end{tabular} \\
    \bottomrule
    \end{NiceTabular}
\end{table*}
\end{landscape}
\subsection{Window size limits}\label{subsec: ws limits}
In order to ensure comparability of the results to \cite{a:Perretti2012} the drift slope estimates are calculated for comparable window sizes and the corresponding $BF_{12,21}$ are calculated to get an impression of the minimal necessary amount of data per window that yields significant results. In \cite{a:Perretti2012} a low-sampled time series variant with one measurement per year and a high-sampled variant of the time series with $50$ data points per year is investigated. Here, we will focus on the high-sampled variants because the discussed indicators including the proposed drift slope are only applicable if the information level in terms of available data is high enough to resolve the considered dynamics. This remains a common limitation of the discussed indicators.\\
However, focusing on the high-sampled datasets with $50$ points per year the $BF_{12,21}$ are calculated for window sizes $\lbrace 150, 100, 50, 25\rbrace$ in decreasing order until the $BF_{12}$ is no longer significant ($BF_{12}\leq 100$). The results for the discussed noise levels and types are summarized in table \ref{table: limit ws without deseasonalization} without deseasonalization and in table \ref{table: limit ws with deseasonalization} with deseasonalization. The color scheme is defined as in subsection \ref{subsec: drift slope analysis}. The tile is signed to be ``inadequate'' if both model evidences are numerically zero. A Bayes factor pair of infinite an zero indicates that one model has an evidence of zero and thus, does not fit the data at all. Without deseasonalization significant results are mainly generated for windows bigger than $50$ and less or equal to $100$ data points except for white noise with $\sigma = 0.1$ where windows less or equal to $50$ data points are sufficient and pink noise with $\sigma = 4.5$ where windows have to be bigger than $100$ data points. Thus, most of the significant windows include a time interval of one up to two years which is mostly comparable to the computations in \cite{a:Perretti2012} assuming windows of one year. Furthermore, a suitable deseasonalization is able to decrease the necessary window size for significant drift slope trends even below one year between more than $25$ and less or equal to $50$ data points for pink and red noise. The performance for small windows tends to become slightly worse for the cases with small and strong white noise $\sigma = \lbrace 0.1,\ 4.5\rbrace$. This is a sign for the difficulties of deseasonalization without removing valuable information for the drift slope estimation at the same time. It has to be mentioned that the drift slope trends for small window sizes as in this limit cases are volatile and thus, less appropriate for an on-line analysis approach.\\
\begin{landscape}
\begin{table*}
\caption{Summary of the Bayes factors comparing a linear model $\mathcal{M}_1$ to a constant model $\mathcal{M}_2$ for the small window sizes of $\lbrace 150, 100, 50 \rbrace$ for various noise types and levels without deseasonalization of the data. Model $\mathcal{M}_1$ is preferred upon the threshold $BF_{12} > 100$ (\cite{b:Jeffreys1998}) colored in green. Grey tiles are insignificant results. The constant model $\mathcal{M}_2$ is never preferred. If both models had an evidence that resulted in a numerical zero, the tile is marked as ``inadequate'', because none of the models was adequate to fit the data. In the case that one evidence was finite and one zero the Bayes factor ratio becomes infinite indicating that the model with an evidence of zero does not fit the data at all and thus, the other one is preferred. The results tend to be significant for more than $50$ and less than or equal to $100$ data points per window except for the red noise with $\sigma = 4.5$ that becomes significant for more than $100$ and less than or equal to $150$ data points. The white noise case with $\sigma =0.1$ is already significant for less than or equal to $50$ data points. The infinite $BF_{12}$ of the pink noise system with noise level $\sigma = 4.5$ is written in brackets and colored in grey, because the trend is very noisy. This corresponds to a period in time between one and two years of high-sampled observation of the ecological system.}
    \label{table: limit ws without deseasonalization}
\ra{1.4}
    \centering
    \begin{NiceTabular}{@{}lrrrrrrrrr}[]\toprule[1.3pt]
     & \multicolumn{3}{c}{white noise ($P_{\rm white} \approx 7.5$)} & \multicolumn{3}{c}{pink noise ($P_{\rm pink} \approx 2 \cdot P_{\rm white}$)} & \multicolumn{3}{c}{red noise ($P_{\rm red} \approx P_{\rm pink}$)} \\
     \cmidrule(lr){2-4} \cmidrule(lr){5-7} \cmidrule(lr){8-10}
    noise level &  $\sigma = 0.1$ & $\sigma = 2.2$ & $\sigma = 4.5$ &  $\sigma = 0.1$ & $\sigma = 2.2$ & $\sigma = 4.5$ &  $\sigma = 0.1$ & $\sigma = 2.2$ & $\sigma = 4.5$ \\ 
    \cmidrule{1 - 10} 
    window size
    & \begin{tabular}{@{}r@{}} $\rm BF_{12}$ \\ $\rm BF_{21}$ \end{tabular} 
    & \begin{tabular}{@{}r@{}} $\rm BF_{12}$ \\ $\rm BF_{21}$ \end{tabular} 
    & \begin{tabular}{@{}r@{}} $\rm BF_{12}$ \\ $\rm BF_{21}$ \end{tabular}
    & \begin{tabular}{@{}r@{}} $\rm BF_{12}$ \\ $\rm BF_{21}$ \end{tabular} 
    & \begin{tabular}{@{}r@{}} $\rm BF_{12}$ \\ $\rm BF_{21}$ \end{tabular} 
    & \begin{tabular}{@{}r@{}} $\rm BF_{12}$ \\ $\rm BF_{21}$ \end{tabular} 
    & \begin{tabular}{@{}r@{}} $\rm BF_{12}$ \\ $\rm BF_{21}$ \end{tabular}
    & \begin{tabular}{@{}r@{}} $\rm BF_{12}$ \\ $\rm BF_{21}$ \end{tabular}
    & \begin{tabular}{@{}r@{}} $\rm BF_{12}$ \\ $\rm BF_{21}$ \end{tabular}
     \\ 
    \midrule
    $50$
    & \cellcolor{green}\begin{tabular}{@{}r@{}} $2.6 \cdot 10^{5}$ \\ $3.9 \cdot 10^{-6}$ \end{tabular}
    & \cellcolor{gray}\begin{tabular}{@{}r@{}} inadequate \end{tabular}
    & \cellcolor{gray}\begin{tabular}{@{}r@{}} inadequate \end{tabular}
    & \cellcolor{gray}\begin{tabular}{@{}r@{}} $1.6\cdot 10^{-2}$ \\ $63$ \end{tabular}
    & \cellcolor{gray}\begin{tabular}{@{}r@{}} $1.2\cdot 10^{-2}$ \\ $84$ \end{tabular}
    & \cellcolor{gray}\begin{tabular}{@{}r@{}} $(\infty)$ \\ (0) \end{tabular}
   & \cellcolor{gray}\begin{tabular}{@{}r@{}} $79$ \\ $1.3\cdot 10^{-2}$ \end{tabular}
    & \cellcolor{gray}\begin{tabular}{@{}r@{}} $1.2$ \\ $0.85$ \end{tabular}
    & \cellcolor{gray}\begin{tabular}{@{}r@{}} inadequate \end{tabular} \\
    $100$
    & \cellcolor{green}\begin{tabular}{@{}r@{}} $\infty$ \\ 0 \end{tabular}
    & \cellcolor{green}\begin{tabular}{@{}r@{}} $\infty$ \\ 0 \end{tabular}
    & \cellcolor{green}\begin{tabular}{@{}r@{}} $\infty$ \\ 0 \end{tabular}
    & \cellcolor{green}\begin{tabular}{@{}r@{}} $\infty$ \\ 0 \end{tabular}
    & \cellcolor{green}\begin{tabular}{@{}r@{}} $\infty$ \\ 0 \end{tabular}
    & \cellcolor{green}\begin{tabular}{@{}r@{}} $\infty$ \\ 0 \end{tabular}
    & \cellcolor{green}\begin{tabular}{@{}r@{}} $\infty$ \\ 0 \end{tabular}
    & \cellcolor{green}\begin{tabular}{@{}r@{}} $\infty$ \\ 0 \end{tabular}
    & \cellcolor{gray}\begin{tabular}{@{}r@{}} inadequate \end{tabular} \\
    $150$
    & \cellcolor{green}\begin{tabular}{@{}r@{}} $\infty$ \\ 0 \end{tabular}
    & \cellcolor{green}\begin{tabular}{@{}r@{}} $\infty$ \\ 0 \end{tabular}
    & \cellcolor{green}\begin{tabular}{@{}r@{}} $\infty$ \\ 0 \end{tabular}
    & \cellcolor{green}\begin{tabular}{@{}r@{}} $\infty$ \\ 0 \end{tabular}
    & \cellcolor{green}\begin{tabular}{@{}r@{}} $\infty$ \\ 0 \end{tabular}
    & \cellcolor{green}\begin{tabular}{@{}r@{}} $\infty$ \\ 0 \end{tabular}
    & \cellcolor{green}\begin{tabular}{@{}r@{}} $\infty$ \\ 0 \end{tabular}
    & \cellcolor{green}\begin{tabular}{@{}r@{}} $\infty$ \\ 0 \end{tabular}
    & \cellcolor{green}\begin{tabular}{@{}r@{}} $\infty$ \\ 0 \end{tabular} \\
\bottomrule
\end{NiceTabular}
\end{table*}
\end{landscape} 
\vspace{25pt}
\begin{landscape}
\begin{table*}
\caption{Same as table \ref{table: limit ws without deseasonalization} for window sizes of $\lbrace 150, 100, 50, 25 \rbrace$ with deseasonalization of the data. A deseasonalization deceases the necessary data per window to generate significant results to less than or equal to $50$ and more than $25$ data points for all pink and red noise cases. This corresponds to time periods of a half year up to one year of observation in a high-sampled manner. The slightly worse results for the white noise cases give a hint that the method reacts sensitive to the deseasonalization in that noise case.}
    \label{table: limit ws with deseasonalization}
\ra{1.4}
    \centering
    \begin{NiceTabular}{@{}lrrrrrrrrr}[]\toprule[1.3pt]
     & \multicolumn{3}{c}{white noise ($P_{\rm white} \approx 7.5$)} & \multicolumn{3}{c}{pink noise ($P_{\rm pink} \approx 2 \cdot P_{\rm white}$)} & \multicolumn{3}{c}{red noise ($P_{\rm red} \approx P_{\rm pink}$)} \\
     \cmidrule(lr){2-4} \cmidrule(lr){5-7} \cmidrule(lr){8-10}
    noise level &  $\sigma = 0.1$ & $\sigma = 2.2$ & $\sigma = 4.5$ &  $\sigma = 0.1$ & $\sigma = 2.2$ & $\sigma = 4.5$ &  $\sigma = 0.1$ & $\sigma = 2.2$ & $\sigma = 4.5$ \\ 
    \cmidrule{1 - 10} 
    window size
    & \begin{tabular}{@{}r@{}} $\rm BF_{12}$ \\ $\rm BF_{21}$ \end{tabular} 
    & \begin{tabular}{@{}r@{}} $\rm BF_{12}$ \\ $\rm BF_{21}$ \end{tabular} 
    & \begin{tabular}{@{}r@{}} $\rm BF_{12}$ \\ $\rm BF_{21}$ \end{tabular}
    & \begin{tabular}{@{}r@{}} $\rm BF_{12}$ \\ $\rm BF_{21}$ \end{tabular} 
    & \begin{tabular}{@{}r@{}} $\rm BF_{12}$ \\ $\rm BF_{21}$ \end{tabular} 
    & \begin{tabular}{@{}r@{}} $\rm BF_{12}$ \\ $\rm BF_{21}$ \end{tabular} 
    & \begin{tabular}{@{}r@{}} $\rm BF_{12}$ \\ $\rm BF_{21}$ \end{tabular}
    & \begin{tabular}{@{}r@{}} $\rm BF_{12}$ \\ $\rm BF_{21}$ \end{tabular}
    & \begin{tabular}{@{}r@{}} $\rm BF_{12}$ \\ $\rm BF_{21}$ \end{tabular}
     \\ 
    \midrule
    $25$
    & \cellcolor{gray}\begin{tabular}{@{}r@{}} - \\ - \end{tabular}
    & \cellcolor{gray}\begin{tabular}{@{}r@{}} - \\ - \end{tabular}
    & \cellcolor{gray}\begin{tabular}{@{}r@{}} - \\ - \end{tabular}
    & \cellcolor{gray}\begin{tabular}{@{}r@{}} inadequate \end{tabular}
    & \cellcolor{gray}\begin{tabular}{@{}r@{}} inadequate  \end{tabular}
    & \cellcolor{gray}\begin{tabular}{@{}r@{}} inadequate \end{tabular}
    & \cellcolor{gray}\begin{tabular}{@{}r@{}} inadequate \end{tabular}
    & \cellcolor{gray}\begin{tabular}{@{}r@{}} inadequate  \end{tabular}
    & \cellcolor{gray}\begin{tabular}{@{}r@{}} inadequate \end{tabular}\\
    $50$
    & \cellcolor{gray}\begin{tabular}{@{}r@{}} inadequate  \end{tabular}
    & \cellcolor{gray}\begin{tabular}{@{}r@{}} inadequate \end{tabular}
    & \cellcolor{gray}\begin{tabular}{@{}r@{}} inadequate  \end{tabular}
    & \cellcolor{green}\begin{tabular}{@{}r@{}} $\infty$ \\ 0 \end{tabular}
    & \cellcolor{green}\begin{tabular}{@{}r@{}} $\infty$ \\ 0 \end{tabular}
    & \cellcolor{green}\begin{tabular}{@{}r@{}} $\infty$ \\ 0 \end{tabular}
    & \cellcolor{green}\begin{tabular}{@{}r@{}} $\infty$ \\ 0 \end{tabular}
    & \cellcolor{green}\begin{tabular}{@{}r@{}} $\infty$ \\ 0 \end{tabular}
    & \cellcolor{green}\begin{tabular}{@{}r@{}} $\infty$ \\ 0 \end{tabular} \\
    $100$
    & \cellcolor{green}\begin{tabular}{@{}r@{}} $\infty$ \\ 0 \end{tabular}
    & \cellcolor{green}\begin{tabular}{@{}r@{}} $\infty$ \\ 0 \end{tabular}
    & \cellcolor{gray}\begin{tabular}{@{}r@{}} inadequate \end{tabular}
    & \cellcolor{green}\begin{tabular}{@{}r@{}} $\infty$ \\ 0 \end{tabular}
    & \cellcolor{green}\begin{tabular}{@{}r@{}} $\infty$ \\ 0 \end{tabular}
    & \cellcolor{green}\begin{tabular}{@{}r@{}} $\infty$ \\ 0 \end{tabular}
    & \cellcolor{green}\begin{tabular}{@{}r@{}} $\infty$ \\ 0 \end{tabular}
    & \cellcolor{green}\begin{tabular}{@{}r@{}} $\infty$ \\ 0 \end{tabular}
    & \cellcolor{green}\begin{tabular}{@{}r@{}} $\infty$ \\ 0 \end{tabular} \\
    $150$
    & \cellcolor{green}\begin{tabular}{@{}r@{}} $\infty$ \\ 0 \end{tabular}
    & \cellcolor{green}\begin{tabular}{@{}r@{}} $\infty$ \\ 0 \end{tabular}
    & \cellcolor{green}\begin{tabular}{@{}r@{}} $\infty$ \\ 0 \end{tabular}
    & \cellcolor{green}\begin{tabular}{@{}r@{}} $\infty$ \\ 0 \end{tabular}
    & \cellcolor{green}\begin{tabular}{@{}r@{}} $\infty$ \\ 0 \end{tabular}
    & \cellcolor{green}\begin{tabular}{@{}r@{}} $\infty$ \\ 0 \end{tabular}
    & \cellcolor{green}\begin{tabular}{@{}r@{}} $\infty$ \\ 0 \end{tabular}
    & \cellcolor{green}\begin{tabular}{@{}r@{}} $\infty$ \\ 0 \end{tabular}
    & \cellcolor{green}\begin{tabular}{@{}r@{}} $\infty$ \\ 0 \end{tabular} \\
    \bottomrule
    \end{NiceTabular}
\end{table*}
\end{landscape}
\section{Summary and conclusion} \label{sec: conclusion}
Our investigations are based on the destabilizing ecological model previously considered in \cite{a:Perretti2012} with white, correlated and weak up to strong noise geared to real world experimental data. The simulations are almost comparable except for a slightly longer period of data sampling before the ``point of no return''.\\
The main difficulties stated in \cite{a:Perretti2012} concerning the applicability of established leading indicator candidates as AR1, std $\hat{\sigma}$, skewness $\gamma$ and kurtosis $\omega$ are given by the conditions of ecological data acquisition: Normally, just short time series with a low sampling rate and strong noise are available. Furthermore, the systems tend to be influenced by correlated pink or red noise and seasonality. The above mentioned early warning signals fail under these circumstances especially due to low data availability for their estimation and high noise levels. Besides, even under favourable simulation conditions the leading indicator candidates are not as reliable as necessary for management decisions (\cite{a:Biggs2009}). In the course of this work we have introduced an alternative leading indicator, the so-called ``drift slope'', and evaluated its performance in comparison to the common leading indicators mentioned above. The drift slope is derived from the MCMC-estimated parameters of the drift term of a stochastic differential equation while the drift term is approximated by a third-order Taylor polynomial.\\
We could show that the drift slope gives reliable trends to estimate the resilience of the system almost regardless of the noise level and type and it fulfills the demands for an early warning signal stated by \cite{a:Biggs2009} which we cite in section \ref{sec: introduction}: The drift slope 
\begin{enumerate}
    \item exhibits a clear threshold of destabilization at zero and the relative distance to zero measures the level of resilience,
    \item provides trends which are easy-to-interpret regarding the necessity of management action,
    \item is comparable across systems in similar contexts because of its parametric ansatz and quantitative nature.
\end{enumerate}
The standard measures skewness $\gamma$ and kurtosis $\omega$ turn out to usually fail to predict the destabilization process which coincides with the observations in \cite{a:Perretti2012}. The kurtosis $\omega$ exhibits non-monotone or ambiguous behaviour and is not suited to be applied as leading indicator in this study. Without deseasonalization the skewness $\gamma$ shows only fragmentary significant results and thus, is not reliable over the range of the considered cases. With deseasonalization the skewness $\gamma$ yields at least significant results under correlated noise conditions. In contrast to the results of \cite{a:Perretti2012} the std $\hat{\sigma}$ also fails to generate significant results whereas the AR1 seems to be the most robust of the standard measures. Nevertheless, the AR1 is very sensitive to the seasonality of the time series that seems to play an important role in the calculations of the leading indicators in general. Deseasonalization has to be taken into account to achieve optimal results, if the noise intensity does not hide the seasonal component. Accordingly, the applicability of the AR1 is enlarged to correlated situations and the clearness of the drift slope trends could be improved. Furthermore, the minimum of necessary data per window for the drift slope estimation could be diminished due to a deseasonalization of the time series. The minimum of available data for the pink and red noise cases is decreased from between $50$ and $100$ to $25-50$ data points except for the red noise case with $\sigma = 4.5$ and thus lie in the observation range of one year or less. The white noise cases do not benefit in that way from a deseasonalization.\\
We considered the destabilization due to a bifurcation, but in principle the Langevin estimation lends itself to monitor changes of the noise level at the same time which can be crucial for systems with the threat of noise-induced transitions (cf. figure 5 (\cite{a:Hessler2021})).\\
In the end, the drift slope could be an interesting alternative in order to deal with very noisy correlated data under typical circumstances in ecology and other fields, but it is limited due to the available amount of data. The low-sampled scenarios with one point per year are impossible to handle neither with the drift slope estimation nor with the standard measures. However, in some cases the opportunities of tracking resilience with the drift slope measure might be an attractive reason to improve sampling-rates and data collection e.g. by using deep-learning image recognition techniques (\cite{a:Francisco2020, a:Duporge2020, a:Zhao2020, a:Aspillaga2021}) for experimental and management purposes, wherever possible.\\

\subsection*{Data and software availability}
The simulated data and Python codes are available on github via \url{https://github.com/MartinHessler/Quantifying_resilience_under_realistic_noise} under a \textit{GNU General Public License v3.0}. The open source python-implementation of the described methods is named \textit{antiCPy} and can be found at \url{https://github.com/MartinHessler/antiCPy} under a \textit{GNU General Public License v3.0}.\\
\subsection*{Acknowledgements}
M. H. thanks the Studienstiftung des deutschen Volkes for a scholarship including financial support. We thank colleagues and friends for proofreading the manuscript.

\bibliographystyle{unsrtnat}
\bibliography{reference}

\end{document}